\documentclass[runningheads]{llncs}
\usepackage{graphicx}
\usepackage[table ]{ xcolor}
\usepackage{booktabs}
\usepackage{epsfig}
\usepackage{amsmath}
\usepackage{amssymb}
\usepackage{amsmath,bm}
\usepackage{multirow}
\usepackage{multicol}
\usepackage{bbding}
\begin{document}
\title{ACN: Adversarial Co-training Network for Brain Tumor Segmentation with Missing Modalities}
\titlerunning{ACN: Adversarial Co-training Network}
%
\author{Yixin Wang\inst{1,2} \and
Yang Zhang\inst{3}\inst{(}\Envelope\inst{)}\and
Yang Liu\inst{1,2} \and
Zihao Lin\inst{4}\and
Jiang Tian\inst{3}\and
Cheng Zhong\inst{3}\and
Zhongchao Shi\inst{3}\and
Jianping Fan\inst{3,5}\and
Zhiqiang He\inst{1,2,3}\inst{(}\Envelope\inst{)}}
\authorrunning{Y. Wang et al.}
%
\institute{Institute of Computing Technology, Chinese Academy of Sciences, Beijing, China  \and
University of Chinese Academy of Sciences, Beijing, China
\email{wangyixin19@mails.ucas.ac.cn}\\ \and
Lenovo Ltd., Beijing, China \and
Department of Electronic \& Computer Engineering, Duke University, NC, USA \and
Department of Computer Science, University of North Charlotte, NC, USA
}
\renewcommand{\thefootnote}{}
\maketitle              
\footnote{This work was done at AI Lab, Lenovo Research.}
\begin{abstract}
   Accurate segmentation of brain tumors from magnetic resonance imaging (MRI) is clinically relevant in diagnoses, prognoses and surgery treatment, which requires multiple modalities to provide complementary morphological and physiopathologic information. However, missing modality commonly occurs due to image corruption, artifacts, different acquisition protocols or allergies to certain contrast agents in clinical practice. Though existing efforts demonstrate the possibility of a unified model for all missing situations, most of them perform poorly when more than one modality is missing. In this paper, we propose a novel Adversarial Co-training Network (ACN) to solve this issue, in which a series of independent yet related models are trained dedicated to each missing situation with significantly better results. Specifically, ACN adopts a novel co-training network, which enables a coupled learning process for both full modality and missing modality to supplement each other's domain and feature representations, and more importantly, to recover the `missing' information of absent modalities. Then, two unsupervised modules, i.e., entropy and knowledge adversarial learning modules are proposed to minimize the domain gap while enhancing prediction reliability and encouraging the alignment of latent representations, respectively. We also adapt modality-mutual information knowledge transfer learning to ACN to retain the rich mutual information among modalities. Extensive experiments on BraTS2018 dataset show that our proposed method significantly outperforms all state-of-the-art methods under any missing situation.
\end{abstract}
\section{Introduction}
Malignant brain tumors have become an aggressive and dangerous disease that leads to death worldwide. Accurate segmentation of brain tumor is crucial to quantitative assessment of tumor progression and surgery treatment planning. Magnetic resonance imaging (MRI) provides various tissue contrast views and spatial resolutions for brain examination \cite{dataset2,yan2020neural}, by which tumor regions can be manually segmented into heterogeneous subregions (i.e., GD-enhancing tumor, peritumoral edema, and the necrotic and non-enhancing tumor core) by comparing MRI modalities with different contrast levels (i.e., T1, T1ce, T2 and Flair). These modalities are essential as they provide each other with complementary information about brain structure and physiopathology.
\begin{figure}
\includegraphics[scale=0.21]{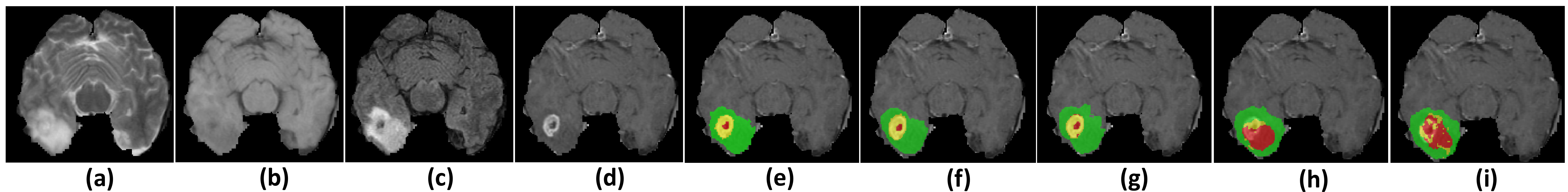}
\caption{Example images. (a)-(d) Four modalities in brain MRI images; (e) Ground-truth of three brain tumors: enhancing tumor (yellow), edema (green), necrotic and non-enhancing tumor (red); (f) Our proposed ACN method; (g) KD-Net \cite{Hu2020KnowledgeDF}; (h) HeMIS\cite{HeMIS}; (i) U-HVED \cite{Reuben}.}
\label{intro}
\end{figure}
In recent years, deep learning has been widely adopted in medical image analysis due to its promising performance as well as the high cost of manual diagnosis. Joint learning from multiple modalities significantly boosts the segmentation or registration accuracy, and thus has been successfully applied in previous works \cite{HAVAEI201718,ZHOU2019100004,wang2020modalitypairing,Kamnitsas,ZheXu01,ZheXu02}.
In clinical settings, however, MRI sequences acquired may be partially unusable or missing due to corruption, artifacts, incorrect machine settings, allergies to certain contrast agents or limited scan time. In such a situation,in only a subset of the full treatment modalities is available.

Despite efforts made for missing modalities, existing works usually learn common representations from full modality, and build a uniform model for all possible missing situations during inference \cite{Tulder,GAN_missing,Chartsias,HeMIS,Chen,Reuben}. However, they perform much worse when more than one modality is missing, which is quite common in clinical practice. Therefore, instead of a `catch-all' general model for all missing situations with low accuracies, it is more valuable to train a series of independent yet related models dedicated to each missing situation with good accuracy.
Hu et al. \cite{Hu2020KnowledgeDF} propose a `dedicated' knowledge distillation method from multi-modal to mono-modal segmentation. However, bias easily occurs to training a single model for a single modality. Moreover, their models are unable to learn domain knowledge from full modality and rich features from different levels. What's more, the teacher model may contain information irrelevant to the mono-modal model, which wastes the network's learning capacity.

To address the obstacles mentioned above, we propose a novel Adversarial Co-training Network (ACN) for missing modalities, which enables a coupled learning process from both full modality and missing modality to supplement each other's domain and feature representations, and more importantly, to recover the `missing' information. Specifically, our model consists of two learning paths: a multimodal path to derive rich modality information and a unimodal path to generate modality-specific representations. Then, a co-training approach is introduced to establish a coupled learning process between them, which is achieved by three major components, namely, 1) an entropy adversarial learning module (EnA) to bridge the domain gap between multimodal and unimodal path; 2) a knowledge adversarial learning module (KnA) to encourage the alignment of latent representations during multimodal and unimodal learning; 3) a modality-mutual information knowledge transfer module (MMI) to recover the most relevant knowledge for incomplete modalities from different feature levels via variational information maximization. The proposed ACN outperforms all state-of-the-arts on a popular brain tumor dataset. Fig. \ref{intro} highlights our superior results under the missing condition that only one modality is avaliable. To the best of our knowledge, this is the \textbf{first} attempt to introduce the concept of unsupervised domain adaptation (UDA) to missing modalities and the \textbf{first} exploration of transferring knowledge from full modality to missing modality from the perspective of information theory. 
\section{Method}
\subsection{ACN: Adversarial Co-training Network}
\begin{figure*}
\begin{center}
\includegraphics[scale=0.20]{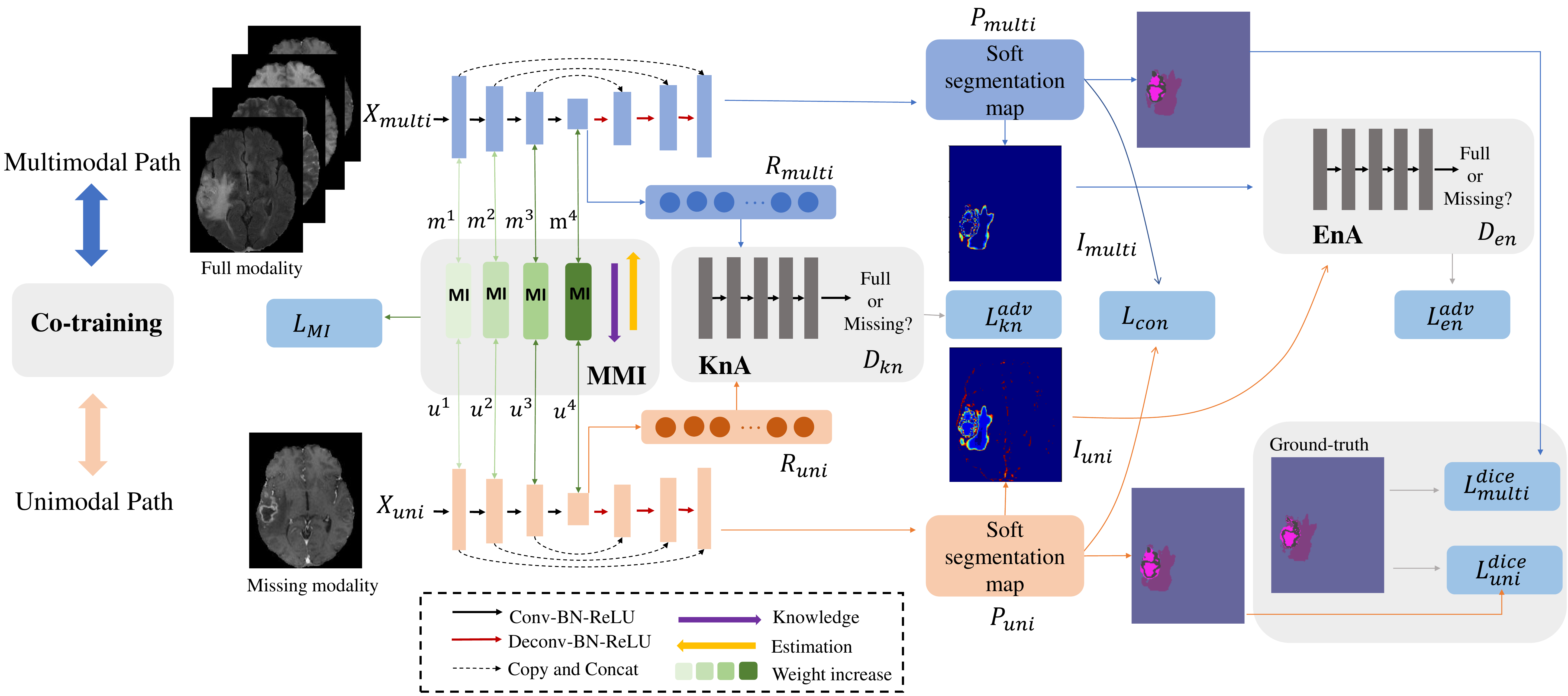}
\end{center}
   \caption{Overview of our proposed ACN, consisting of a multimodal path, a unimodal path and three modules: (a) an entropy adversarial learning module (EnA); (b) a knowledge adversarial learning module (KnA); (c) a modality-mutual information knowledge transfer module (MMI). In (a), given full modality and missing modality, $D_{en}$ is trained to predict domain label based on entropy map $\{I_{multi}, I_{uni}\}$. In (b), $D_{kn}$ is trained to discriminate high-level features $\left\{R_{multi}, R_{uni}\right\}$ to encourage soft-alignment of knowledge. In (c), given $K$ pairs of representations $\left\{\left(m^{(k)}, u^{(k)}\right)\right\}_{k=1}^{K}$ from both paths' encoder layers, `missing' knowledge from multiple levels is recovered via variational information maximization.}
\label{overview}
\end{figure*}

Co-training approach \cite{Co-training} has been applied to tasks with multiple feature views. Each view of the data is independently trained and provides independent and complementary information. Similarly, the fusion of all brain MRI modalities provides complete anatomical and functional information about brain structure and physiopathology. Models trained on incomplete modalities can still focus on the specific tumor or structure of brain MRI and provide important complementary information.
Therefore, we propose ACN (see Fig. \ref{overview}) to enable a coupled learning process to enhance the learning ability of both multimodal and unimodal training. It consists of two learning paths, i.e., the multimodal path and the unimodal path, which are responsible for deriving generic features from full modality and the most relevant features from available incomplete modalities, respectively.
Concretely, patches from full modality are concatenated together to generate an input $X_{multi} \in \mathbb{R}^{H \times W \times M}$ to the multimodal path, where $M=4$ (Flair, T2, T1 and T1ce) in our task.
The unimodal path receives patches from N available incomplete modalities, denoted by $X_{uni} \in \mathbb{R}^{H \times W \times N}$. 
The two paths share the same U-Net architecture and are trained independently and simultaneously. We distill semantic knowledge from the output distributions of both paths and design a consistency loss term $\mathcal{L}_{\mathrm{con}}$ to minimize their Kullback-Leibler (KL) divergence, which is defined as:
\begin{equation}
\!\!\mathcal{L}_{\mathrm{con}}\!\!
=\!\!\frac{1}{C}\!\sum_{c}\!\left(\!\mathcal{D}_{\mathrm{KL}}\!\left(\mathbf{s}_{c}^{m}\| \mathbf{s}_{c}^{u}\!\right)\!\!+\!\!\mathcal{D}_{\mathrm{KL}}\!\left(\!\mathbf{s}_{c}^{u}\| \mathbf{s}_{c}^{m}\!\right)\!\right)\!\!=\!\!\frac{1}{C}\!\!\sum_{c}\!\!\left(\!\sum\! \mathbf{s}_{c}^{m}\!\log\!\frac{\mathbf{s}_{c}^{m}}{\mathbf{s}_{c}^{u}}\!\!+\!\!\sum\!\mathbf{s}_{c}^{u}\!\log\! \frac{\mathbf{s}_{c}^{u}}{\mathbf{s}_{c}^{m}}\!\right)\!,\!
\end{equation}
where $\mathbf{s}_{c}^{m}$ and $\mathbf{s}_{c}^{u}$ denote the softened logits of multimodal and unimodal path. $C$ is the total
number of classes. 
Such a coupled learning process not only promotes the unimodal path to learn from multimodal path but also adds necessary guidance and regularization to multimodal path to further replenish the learned generic features.

This consistency exploits the correlations between full modality and missing modality in the output level. However, the two paths learn from different modalities and domains. It is significant to explicitly enhance the alignment of different domains and high-level knowledge. Therefore, we propose two adversarial learning modules to minimize the distribution distance between the two paths and utilize knowledge transfer to retain rich mutual information from a perspective of information theory.
\subsection{Entropy Adversarial Learning}
The prediction of multimodal path is not necessarily more accurate than unimodal path. However, models trained on multiple modalities tend to be more reliable than those on incomplete modalities. Therefore, we adopt the principle of entropy maps as a confidence measurement from unsupervised domain adaptation (UDA) tasks \cite{Vu_2019_CVPR,IntraDA} and design an entropy adversarial module to match the distributions between two paths.

In detail, the multimodal and unimodal path receive $X_{multi}$ and $X_{uni}$ separately, along with their corresponding pixel-level C-class ground-truth $Y^{(h,w)} = \left[Y^{(h, w, c)}\right]_{c}$. The multimodal path takes $X_{multi}$ as input and generates the soft segmentation map $\left[P_{multi}^{(h, w, c)}\right]_{c}$ at the output level, which represents the discrete distribution over $C$ classes. It is observed that predictions from unimodal path tend to be under-confident with high-entropy. Conversely, predictions from multimodal path are usually over-confident with low-entropy. Therefore, we adopt a unified adversarial training strategy to minimize the entropy of unimodal learning by encouraging its entropy distribution to be more similar to the multimodal one. Concretely, the entropy map $I_{multi}$ of multimodal path is defined in the same way as weighted self-information maps \cite{Vu_2019_CVPR}, calculated by:
\begin{equation}
I_{multi}^{(h, w)}=\sum_{c}-P_{multi}^{(h, w, c)} \cdot \log P_{multi}^{(h, w, c)}.
\end{equation}
Similarly, receiving missing modality patches $X_{uni}$, the unimodal path can be seen as a generator $G_{en}$, which generates the soft segmentation map $\left[P_{uni}^{(h, w, c)}\right]_{c}$ and entropy map $I_{uni}$. These pixel-level vectors can be seen as the disentanglement of the Shannon Entropy, which reveals the prediction confidence. We then introduce a fully-convolutional network as a discriminator $D_{en}$. The generator $G_{en}$ tries to generate an unimodal entropy map $I_{uni}$ and fools $D_{en}$, while the discriminator $D_{en}$ aims to distinguish $I_{uni}$ from the multimodal entropy map $I_{multi}$. Accordingly, the optimization of $G_{en}$ and $D_{en}$ is achieved by the following objective function:
\begin{equation}
\begin{array}{r}
\mathcal{L}_{en}^{adv}\left(X_{multi}, X_{uni}\right)=\sum\limits_{h, w} \log \left(1-D_{en}\left(I_{uni}^{(h, w)}\right)\right)+\log \left(D_{en}\left(I_{multi}^{(h, w)}\right)\right).
\end{array}
\end{equation}
\subsection{Knowledge Adversarial Learning}
Considering that the high-level representations contain richer information, we also need to encourage features distribution alignment in latent space. Simply minimizing the KL divergence between the features of two paths' bottlenecks $\left\{R_{multi}, R_{uni}\right\}$ may easily disturb the underlying learning of unimodal path in the deep layers. Therefore, we encourage the high-level representations of both paths to be aligned using a knowledge adversarial module.
Similar to EnA, the generator $G_{kn}$ tries to generate high-level features to mislead another discriminator $D_{kn}$. The objective function of this process is formulated as:
\begin{equation}
\begin{array}{r}
\mathcal{L}_{kn}^{adv}\left(X_{multi}, X_{uni}\right)= \log \left(1-D_{kn}\left(R_{uni}\right)\right) +\log \left(D_{kn}\left(R_{multi}\right)\right).
\end{array}
\end{equation}
KnA serves as a soft-alignment to encourage unimodal path to learn abundant and `missing' knowledge from full modality.
\subsection{Modality-mutual Information Knowledge Transfer Learning}
Multimodal path may contain information irrelevant to the task, which requires superfluous effort for alignment by unimodal path. To address this issue, we introduce the modality-mutual information knowledge transfer learning to retain high mutual information between two paths. Specifically, $K$ pairs of representations $\left\{\left(m^{(k)}, u^{(k)}\right)\right\}_{k=1}^{K}$ can be obtained from $K$ encoder layers of the multimodal and unimodal path separately. Given the entropy $H(m)$ and conditional entropy $H(m\mid u)$, the mutual information $MI$ between each pair $(m,u)$ can be defined by: $MI(m, u)=H(m)-H(m\mid u)$, which measures a reduction in uncertainty in the knowledge of the multimodal learning encoded in its layers when the unimodal knowledge is known. 
Following \cite{ahn2019variational} to measure the exact values, we use variational information maximization \cite{VIM} for each $MI(m, u)$:
\begin{equation}
\begin{aligned}
MI(m, u) &=H(m)\!-\!H(m\!\mid\!u) =H(m)\!+\!\mathbb{E}_{m, u \sim p(m, u)}[\log p(m\!\mid\!u)] \\
&=H(m)\!+\!\mathbb{E}_{u \sim p(u)}\!\left[D_{\mathrm{KL}}(p(m\!\mid\!u) \| q(m\!\mid\!u))\right]\!+\!\mathbb{E}_{m, u \sim p(m, u)}\!\left[\log q\left(m\!\mid\!u\right)\right] \\
& \geq H(m)\!+\!\mathbb{E}_{m, u \sim p(m, u)}[\log q(m\!\mid\!u)].
\end{aligned}
\end{equation}
The distribution $p(m \mid u)$ is approximated by a variational distribution $q(m \mid u)$. Accordingly, the optimization of hierarchical mutual information can be formulated with the following loss function $\mathcal{L_{MI}}$:
\begin{equation}
\mathcal{L_{MI}}\!=\!-\!\!\sum_{k=1}^{K}\!\gamma_{k} M\!I\!\left(\!\!m^{(k)}\!,\!u^{(k)}\!\right)\!\!=\!-\!\!\sum_{k=1}^{K}\! \gamma_{k}\mathbb{E}_{m^{(k)}\!, u^{(k)} \sim p\left(\!m^{(k)}\!\!,u^{(k)}\!\right)}\!\!\left[\log q\!\left(\!\!m^{(k)}\!\!\mid \!u^{(k)}\!\right)\!\right]\!,\!
\end{equation}
where $\gamma_{k}$ increases with $k$, indicating that higher layers contain more semantic information which should be assigned with larger weights for guidance. For specific implementation of our co-training network, the variation distribution can be realized by: 
\begin{equation}
\!-\log q(m\!\mid\!u)\!=\!\sum_{c=1}^{C}\!\sum_{h=1}^{H}\!\sum_{w=1}^{W}\!\log \sigma_{c}+\frac{\left(\!m_{c, h, w}\!-\!\mu_{c, h, w}(u)\!\right)^{2}}{2\sigma_{c}^{2}}+\text {Z},
\end{equation}
where $\mu(\cdot)$ and $\sigma$ denote the heteroscedastic mean and homoscedastic variance function of a Gaussian distribution. $W$ and $H$ denote width and height, C is the channel numbers of the corresponding layers and $\text {Z}$ is a constant value.

\subsection{Overall Loss and Training Procedure}
The overall loss funtion $\mathcal{L}$ is formulated by:
\begin{equation}
\begin{aligned}
\mathcal{L} =&\lambda_{multi} \mathcal{L}_{multi}^{dice} + \lambda_{uni} \mathcal{L}_{uni}^{dice} + \omega(t) \mathcal{L}_{con} + \lambda_{0}\mathcal{L}_{en}^{adv} + \lambda_{1}\mathcal{L}_{kn}^{adv} +\lambda_{2}\mathcal{L}_{MI}.
\end{aligned}
\end{equation}
where $\mathcal{L}_{multi}^{dice}$ and $\mathcal{L}_{uni}^{dice}$ are commonly used segmentation Dice loss in medical tasks for multimodal and unimodal path respectively. $\omega(t)=0.1 * e^{\left(-5\left(1-S / L\right)^{2}\right)}$ is a time-dependent Gaussian weighting function, where $S$ and $L$ represent the current training step and ramp-up length separately. $\lambda_{multi}$, $\lambda_{uni}$, $\lambda_{0}$, $\lambda_{1}$ and $\lambda_{2}$ are trade-off parameters, which are set as 0.2, 0.8, 0.001, 0.0002, and 0.5 in our model. They are chosen according to the performance under the circumstance that only T1ce modality is available since it's the major modality for tumor diagnosis.
The ultimate goal of the overall co-training procedure is the following optimization:$
\underset{G_{en},G_{kn}}{\min } \underset{D_{en},D_{kn}}{\max }  \mathcal{L}$.
\section{Experimental Results}
\subsection{Experimental Setup}
\noindent\textbf{Dataset} 
BraTS2018 training dataset \cite{dataset1,dataset2,dataset3} consists of 285 multi-contrast MRI scans with four modalities: a) native (T1), b) post-contrast T1-weighted (T1ce), c) T2-weighted (T2), and d) T2 Fluid Attenuated Inversion Recovery (Flair) volumes. Each of these modalities captures different properties of brain tumor subregions: GD-enhancing tumor (ET), the peritumoral edema (ED), and the necrotic and non-enhancing tumor core (NCR/NET). These subregions are combined into three nested subregions: whole tumor (WT), tumor core (TC) and enhancing tumor (ET). All the volumes have been co-registered to the same anatomical template and interpolated to the same resolution.\\

\noindent\textbf{Implementation and Evaluation}
Experiments are implemented in Pytorch and performed on NVIDIA Tesla V100 with 32GB Ram. For fair comparison, we follow the experimental settings of the winner of BraTS2018 \cite{18top1}. Both paths share the same U-Net backbone as \cite{18top1} and the discriminators $D_{en}$ and $D_{kn}$ are two fully-convolutional networks. The input patch size is set as $160\times 192\times 128$ and batch size as 1. The adam optimizer is applied, and the initial learning rate is set as $1e-4$ and progressively decreases according to a poly policy $\left(1-\text { epoch } / \text { epoch }_{\max }\right)^{0.9}$, where $\text { epoch }_{\max }$ is the total number of epochs (300).
We randomly split the dataset into training and validation sets by a ratio of 2:1. The segmentation performance is evaluated on each nested subregion of brain tumors using `Dice similarity coefficient (DSC)' and `Hausdorff distance (HD95)'. A higher DSC and a lower HD95 indicate a better segmentation performance. Codes will be available at \url{https://github.com/Wangyixinxin/ACN}.
\subsection{Results and Analysis}
\textbf{Comparisons with State-of-the-art Methods}
We compare the proposed ACN with three state-of-the-art methods: two `catch-all' models HeMIS \cite{HeMIS} and U-HVED \cite{Reuben} and a `dedicated' model KD-Net \cite{Hu2020KnowledgeDF}. For fair comparison, U-Net is applied as the benchmark to all methods.
\begin{table*}[htbp]
  \caption{Comparison with state-of-the-art `catch-all' methods (DSC \%) on three nested subregions (ET, TC and WT). Modalities present are denoted by $\bullet$, the missing ones by $\circ$.}
  \centering
  \resizebox{\textwidth}{37mm}{
    \begin{tabular}{cccccccccccccccccccccc}
    \toprule
    \multicolumn{4}{c}{\multirow{2}[4]{*}{\textbf{Modalities}}} & \multicolumn{6}{c}{\textbf{ET}} & \multicolumn{6}{c}{\textbf{TC}}      & \multicolumn{6}{c}{\textbf{WT}} \\
\cmidrule(lr){5-10} \cmidrule(lr){11-16} \cmidrule(lr){17-22}    \multicolumn{4}{c}{}         & \multicolumn{2}{c}{\textbf{U-HeMIS}} & \multicolumn{2}{c}{\textbf{U-HVED}} & \multicolumn{2}{c}{\textbf{ACN}} & \multicolumn{2}{c}{\textbf{U-HeMIS}} & \multicolumn{2}{c}{\textbf{U-HVED}} & \multicolumn{2}{c}{\textbf{ACN}} & \multicolumn{2}{c}{\textbf{U-HeMIS}} & \multicolumn{2}{c}{\textbf{U-HVED}} & \multicolumn{2}{c}{\textbf{ACN}} \\
    \midrule
    \textbf{Flair} & \textbf{T1} & \textbf{T1ce} & \textbf{T2} & \multicolumn{1}{l}{\cellcolor[rgb]{ .906,  .902,  .902}\textbf{DSC}} & \multicolumn{1}{l}{\textbf{HD95}} & \multicolumn{1}{l}{\cellcolor[rgb]{ .906,  .902,  .902}\textbf{DSC}} & \multicolumn{1}{l}{\textbf{HD95}} & \multicolumn{1}{l}{\cellcolor[rgb]{ .906,  .902,  .902}\textbf{DSC}} & \multicolumn{1}{l}{\textbf{HD95}} & \multicolumn{1}{l}{\cellcolor[rgb]{ .906,  .902,  .902}\textbf{DSC}} & \multicolumn{1}{l}{\textbf{HD95}} & \multicolumn{1}{l}{\cellcolor[rgb]{ .906,  .902,  .902}\textbf{DSC}} & \multicolumn{1}{l}{\textbf{HD95}} & \multicolumn{1}{l}{\cellcolor[rgb]{ .906,  .902,  .902}\textbf{DSC}} & \multicolumn{1}{l}{\textbf{HD95}} & \multicolumn{1}{l}{\cellcolor[rgb]{ .906,  .902,  .902}\textbf{DSC}} & \multicolumn{1}{l}{\textbf{HD95}} & \multicolumn{1}{l}{\cellcolor[rgb]{ .906,  .902,  .902}\textbf{DSC}} & \multicolumn{1}{l}{\textbf{HD95}} & \multicolumn{1}{l}{\cellcolor[rgb]{ .906,  .902,  .902}\textbf{DSC}} & \multicolumn{1}{l}{\textbf{HD95}} \\
    \midrule
     $\circ$     &$\circ$       &$\circ$       & $\bullet$     & \cellcolor[rgb]{ .906,  .902,  .902}25.63  & 14.42  & \cellcolor[rgb]{ .906,  .902,  .902}22.82  & 14.28  & \cellcolor[rgb]{ .906,  .902,  .902}\textbf{42.98 } & \textbf{10.66 } & \cellcolor[rgb]{ .906,  .902,  .902}57.20  & 17.88  & \cellcolor[rgb]{ .906,  .902,  .902}54.67  & 15.38  & \cellcolor[rgb]{ .906,  .902,  .902}\textbf{67.94 } & \textbf{10.07 } & \cellcolor[rgb]{ .906,  .902,  .902}80.96  & 12.53  & \cellcolor[rgb]{ .906,  .902,  .902}79.83  & 14.63  & \cellcolor[rgb]{ .906,  .902,  .902}\textbf{85.55 } & \textbf{7.24 } \\
    \midrule
     $\circ$     &$\circ$       & $\bullet$     &$\circ$       & \cellcolor[rgb]{ .906,  .902,  .902}62.02  & 22.87  & \cellcolor[rgb]{ .906,  .902,  .902}57.64  & 31.90  & \cellcolor[rgb]{ .906,  .902,  .902}\textbf{78.07 } & \textbf{3.57 } & \cellcolor[rgb]{ .906,  .902,  .902}65.29  & 28.21  & \cellcolor[rgb]{ .906,  .902,  .902}59.59  & 38.01  & \cellcolor[rgb]{ .906,  .902,  .902}\textbf{84.18 } & \textbf{5.04 } & \cellcolor[rgb]{ .906,  .902,  .902}61.53  & 28.23  & \cellcolor[rgb]{ .906,  .902,  .902}53.62  & 34.14  & \cellcolor[rgb]{ .906,  .902,  .902}\textbf{80.52} & \textbf{8.42 } \\
    \midrule
     $\circ$     & $\bullet$     &$\circ$       &$\circ$       & \cellcolor[rgb]{ .906,  .902,  .902}10.16  & 26.06  & \cellcolor[rgb]{ .906,  .902,  .902}8.60  & 28.80  & \cellcolor[rgb]{ .906,  .902,  .902}\textbf{41.52 } & \textbf{10.68 } & \cellcolor[rgb]{ .906,  .902,  .902}37.39  & 30.70  & \cellcolor[rgb]{ .906,  .902,  .902}33.90  & 32.29  & \cellcolor[rgb]{ .906,  .902,  .902}\textbf{71.18 } & \textbf{10.46 } & \cellcolor[rgb]{ .906,  .902,  .902}57.62  & 27.40  & \cellcolor[rgb]{ .906,  .902,  .902}49.51  & 31.87  & \cellcolor[rgb]{ .906,  .902,  .902}\textbf{79.34 } & \textbf{10.22 } \\
    \midrule
    $\bullet$     &$\circ$       &$\circ$       &$\circ$       & \cellcolor[rgb]{ .906,  .902,  .902}11.78  & 25.85  & \cellcolor[rgb]{ .906,  .902,  .902}23.80  & 14.39  & \cellcolor[rgb]{ .906,  .902,  .902}\textbf{42.77 } & \textbf{11.44 } & \cellcolor[rgb]{ .906,  .902,  .902}26.06  & 30.69  & \cellcolor[rgb]{ .906,  .902,  .902}57.90  & 15.16  & \cellcolor[rgb]{ .906,  .902,  .902}\textbf{67.72 } & \textbf{11.75 } & \cellcolor[rgb]{ .906,  .902,  .902}52.48  & 28.21  & \cellcolor[rgb]{ .906,  .902,  .902}84.39  & 12.40  & \cellcolor[rgb]{ .906,  .902,  .902}\textbf{87.30 } & \textbf{7.81 } \\
    \midrule
    $\circ$      &$\circ$       & $\bullet$     & $\bullet$     & \cellcolor[rgb]{ .906,  .902,  .902}67.83  & 9.06  & \cellcolor[rgb]{ .906,  .902,  .902}67.83  & 10.70  & \cellcolor[rgb]{ .906,  .902,  .902}\textbf{75.65 } & \textbf{4.36 } & \cellcolor[rgb]{ .906,  .902,  .902}76.64  & 11.16  & \cellcolor[rgb]{ .906,  .902,  .902}73.92  & 14.56  & \cellcolor[rgb]{ .906,  .902,  .902}\textbf{84.41 } & \textbf{6.41 } & \cellcolor[rgb]{ .906,  .902,  .902}82.48  & 10.23  & \cellcolor[rgb]{ .906,  .902,  .902}81.32  & 12.24  & \cellcolor[rgb]{ .906,  .902,  .902}\textbf{86.41 } & \textbf{7.41 } \\
    \midrule
     $\circ$     & $\bullet$     & $\bullet$     &$\circ$       & \cellcolor[rgb]{ .906,  .902,  .902}66.22  & 14.89  & \cellcolor[rgb]{ .906,  .902,  .902}61.11  & 26.92  & \cellcolor[rgb]{ .906,  .902,  .902}\textbf{75.21 } & \textbf{3.77 } & \cellcolor[rgb]{ .906,  .902,  .902}72.46  & 16.86  & \cellcolor[rgb]{ .906,  .902,  .902}67.55  & 29.65  & \cellcolor[rgb]{ .906,  .902,  .902}\textbf{84.59 } & \textbf{5.76 } & \cellcolor[rgb]{ .906,  .902,  .902}68.47  & 19.55  & \cellcolor[rgb]{ .906,  .902,  .902}64.22  & 28.46  & \cellcolor[rgb]{ .906,  .902,  .902}\textbf{80.05 } & \textbf{9.27 } \\
    \midrule
    $\bullet$     & $\bullet$     &$\circ$       &$\circ$       & \cellcolor[rgb]{ .906,  .902,  .902}10.71  & 25.71  & \cellcolor[rgb]{ .906,  .902,  .902}27.96  & 15.09  & \cellcolor[rgb]{ .906,  .902,  .902}\textbf{43.71 } & \textbf{11.38 } & \cellcolor[rgb]{ .906,  .902,  .902}41.12  & 25.06  & \cellcolor[rgb]{ .906,  .902,  .902}61.14  & 14.70  & \cellcolor[rgb]{ .906,  .902,  .902}\textbf{71.30 } & \textbf{11.87 } & \cellcolor[rgb]{ .906,  .902,  .902}64.62  & 20.69  & \cellcolor[rgb]{ .906,  .902,  .902}85.71  & 12.20  & \cellcolor[rgb]{ .906,  .902,  .902}\textbf{87.49 } & \textbf{8.88 } \\
    \midrule
     $\circ$     & $\bullet$     &$\circ$       & $\bullet$     & \cellcolor[rgb]{ .906,  .902,  .902}32.39  & 13.60  & \cellcolor[rgb]{ .906,  .902,  .902}24.29  & 14.33  & \cellcolor[rgb]{ .906,  .902,  .902}\textbf{47.39 } & \textbf{9.10 } & \cellcolor[rgb]{ .906,  .902,  .902}60.92  & 15.18  & \cellcolor[rgb]{ .906,  .902,  .902}56.26  & 14.66  & \cellcolor[rgb]{ .906,  .902,  .902}\textbf{73.28 } & \textbf{8.72 } & \cellcolor[rgb]{ .906,  .902,  .902}82.41  & 11.90  & \cellcolor[rgb]{ .906,  .902,  .902}81.56  & 11.82  & \cellcolor[rgb]{ .906,  .902,  .902}\textbf{85.50 } & \textbf{7.96 } \\
    \midrule
    $\bullet$     &$\circ$       &$\circ$       & $\bullet$     & \cellcolor[rgb]{ .906,  .902,  .902}30.22  & 13.44  & \cellcolor[rgb]{ .906,  .902,  .902}32.31  & 12.84  & \cellcolor[rgb]{ .906,  .902,  .902}\textbf{45.96 } & \textbf{10.45 } & \cellcolor[rgb]{ .906,  .902,  .902}57.68  & 15.73  & \cellcolor[rgb]{ .906,  .902,  .902}62.70  & 12.76  & \cellcolor[rgb]{ .906,  .902,  .902}\textbf{71.61 } & \textbf{10.31 } & \cellcolor[rgb]{ .906,  .902,  .902}82.95  & 11.51  & \cellcolor[rgb]{ .906,  .902,  .902}87.58  & 7.89  & \cellcolor[rgb]{ .906,  .902,  .902}\textbf{87.75 } & \textbf{6.65 } \\
    \midrule
    $\bullet$     &$\circ$       & $\bullet$     &$\circ$       & \cellcolor[rgb]{ .906,  .902,  .902}66.10  & 15.33  & \cellcolor[rgb]{ .906,  .902,  .902}68.36  & 9.66  & \cellcolor[rgb]{ .906,  .902,  .902}\textbf{77.46 } & \textbf{4.22 } & \cellcolor[rgb]{ .906,  .902,  .902}71.49  & 19.34  & \cellcolor[rgb]{ .906,  .902,  .902}75.07  & 11.77  & \cellcolor[rgb]{ .906,  .902,  .902}\textbf{83.35 } & \textbf{5.83 } & \cellcolor[rgb]{ .906,  .902,  .902}68.99  & 19.77  & \cellcolor[rgb]{ .906,  .902,  .902}85.93  & 11.57  & \cellcolor[rgb]{ .906,  .902,  .902}\textbf{88.28 } & \textbf{7.47 } \\
    \midrule
    $\bullet$     & $\bullet$     & $\bullet$     &$\circ$       & \cellcolor[rgb]{ .906,  .902,  .902}68.54  & 12.32  & \cellcolor[rgb]{ .906,  .902,  .902}68.60  & 9.66  & \cellcolor[rgb]{ .906,  .902,  .902}\textbf{76.16} & \textbf{5.62}      & \cellcolor[rgb]{ .906,  .902,  .902}76.01  & 12.43  & \cellcolor[rgb]{ .906,  .902,  .902}77.05  & 10.23  & \cellcolor[rgb]{ .906,  .902,  .902}\textbf{84.25} & \textbf{6.75}      & \cellcolor[rgb]{ .906,  .902,  .902}72.31  & 14.29  & \cellcolor[rgb]{ .906,  .902,  .902}86.72  & 11.11  & \cellcolor[rgb]{ .906,  .902,  .902}\textbf{88.96} &\textbf{6.93}  \\
    \midrule
    $\bullet$     & $\bullet$     &$\circ$       & $\bullet$     & \cellcolor[rgb]{ .906,  .902,  .902}31.07  & 13.94  & \cellcolor[rgb]{ .906,  .902,  .902}32.34  & 11.79  & \cellcolor[rgb]{ .906,  .902,  .902}\textbf{42.09} &\textbf{10.81}       & \cellcolor[rgb]{ .906,  .902,  .902}60.32  & 14.22  & \cellcolor[rgb]{ .906,  .902,  .902}63.14  & 11.58  & \cellcolor[rgb]{ .906,  .902,  .902}\textbf{67.86} & \textbf{10.69}      & \cellcolor[rgb]{ .906,  .902,  .902}83.43  & 11.58  & \cellcolor[rgb]{ .906,  .902,  .902}88.07  & 7.88  & \cellcolor[rgb]{ .906,  .902,  .902}\textbf{88.35} &\textbf{6.14}  \\
    \midrule
    $\bullet$     &$\circ$       & $\bullet$     & $\bullet$     & \cellcolor[rgb]{ .906,  .902,  .902}68.72  & 8.03  & \cellcolor[rgb]{ .906,  .902,  .902}68.93  & 7.72  & \cellcolor[rgb]{ .906,  .902,  .902}\textbf{75.97} & \textbf{4.34}    & \cellcolor[rgb]{ .906,  .902,  .902}77.53  & 9.02  & \cellcolor[rgb]{ .906,  .902,  .902}76.75  & 9.21  & \cellcolor[rgb]{ .906,  .902,  .902}\textbf{82.85} & \textbf{6.48}     & \cellcolor[rgb]{ .906,  .902,  .902}83.85  & 9.26  & \cellcolor[rgb]{ .906,  .902,  .902}88.09  & 8.00  & \cellcolor[rgb]{ .906,  .902,  .902}\textbf{88.34} &\textbf{7.03}  \\
    \midrule
    $\circ$      & $\bullet$     & $\bullet$     & $\bullet$     & \cellcolor[rgb]{ .906,  .902,  .902}69.92  & 7.81  & \cellcolor[rgb]{ .906,  .902,  .902}67.75  & 10.19  & \cellcolor[rgb]{ .906,  .902,  .902}\textbf{76.10} &\textbf{5.01}       & \cellcolor[rgb]{ .906,  .902,  .902}78.96  & 8.93  & \cellcolor[rgb]{ .906,  .902,  .902}75.28  & 10.69  & \cellcolor[rgb]{ .906,  .902,  .902}\textbf{84.67} & \textbf{5.86}      & \cellcolor[rgb]{ .906,  .902,  .902}83.94  & 9.09  & \cellcolor[rgb]{ .906,  .902,  .902}82.32  & 11.08  & \cellcolor[rgb]{ .906,  .902,  .902}\textbf{86.90} & \textbf{6.24} \\
    \midrule
    $\bullet$     & $\bullet$     & $\bullet$     & $\bullet$     & \cellcolor[rgb]{ .906,  .902,  .902}70.24  & 7.43  & \cellcolor[rgb]{ .906,  .902,  .902}69.03  & 8.37  & \cellcolor[rgb]{ .906,  .902,  .902}\textbf{77.06 } & \textbf{5.09 } & \cellcolor[rgb]{ .906,  .902,  .902}79.48  & 8.95  & \cellcolor[rgb]{ .906,  .902,  .902}77.71  & 8.63  & \cellcolor[rgb]{ .906,  .902,  .902}\textbf{85.18 } & \textbf{5.94 } & \cellcolor[rgb]{ .906,  .902,  .902}84.74  & 8.99  & \cellcolor[rgb]{ .906,  .902,  .902}88.46  & 7.80  & \cellcolor[rgb]{ .906,  .902,  .902}\textbf{89.22 } & \textbf{6.71 } \\
    \midrule
    \multicolumn{4}{c}{Average}  & \cellcolor[rgb]{ .906,  .902,  .902}46.10  & 15.38  & \cellcolor[rgb]{ .906,  .902,  .902}46.76  & 15.11  & \cellcolor[rgb]{ .906,  .902,  .902}\textbf{61.21 } & \textbf{7.37 } & \cellcolor[rgb]{ .906,  .902,  .902}62.57  & 17.62  & \cellcolor[rgb]{ .906,  .902,  .902}64.84  & 16.62  & \cellcolor[rgb]{ .906,  .902,  .902}\textbf{77.62 } & \textbf{8.13 } & \cellcolor[rgb]{ .906,  .902,  .902}74.05  & 16.22  & \cellcolor[rgb]{ .906,  .902,  .902}79.16  & 14.87  & \cellcolor[rgb]{ .906,  .902,  .902}\textbf{85.92 } & \textbf{7.62 } \\
    \bottomrule
    \end{tabular}}%
  \label{tab:All_results}%
\end{table*}%
Table \ref{tab:All_results} shows that ACN significantly outperforms both HeMIS and U-HVED for all the 15 possible combinations of missing modality, especially when only one or two modalities are available. It is noted that missing T1ce leads to a severe decreasing on both ET and TC while missing Flair causes a significant drop on WT. Both HeMIS and U-HVED perform poorly under the above circumstances, with unacceptable DSC scores of only about $10\%-20\%$ on the ET and TC. In contrast, our method achieves promising DSC scores of above $40\%$ in such hard situations, which is much more valuable in clinical practice.
Table \ref{compare_kdnet} shows the DSC comparison with the `dedicated' KD-Net. Since KD-Net aims to deal with mono-modal segmentation, we set the comparison under the condition that only one modality is available.
\begin{table}[htbp]
  \centering
    \caption{Comparison with state-of-the-art `dedicated' method KD-Net \cite{Hu2020KnowledgeDF} (DSC \%).}
    \begin{tabular}{cccccccccccc}
    \toprule
    \multicolumn{4}{c}{\textbf{Modalities}} & \multicolumn{2}{c}{\textbf{ET}} & \multicolumn{2}{c}{\textbf{TC}} & \multicolumn{2}{c}{\textbf{WT}} & \multicolumn{2}{c}{\textbf{Average}} \\
    \cmidrule{1-4} \cmidrule(lr){5-6} \cmidrule(lr){7-8} \cmidrule(lr){9-10} \cmidrule(lr){11-12}
    \textbf{Flair} & \multicolumn{1}{c}{\textbf{T1}} & \multicolumn{1}{c}{\textbf{T1ce}} & \multicolumn{1}{c}{\textbf{T2}} & \multicolumn{1}{c}{\textbf{KD-Net}} & \multicolumn{1}{c}{\cellcolor[rgb]{ .906,  .902,  .902}\textbf{ACN}} & \multicolumn{1}{c}{\textbf{KD-Net}} & \multicolumn{1}{c}{\cellcolor[rgb]{ .906,  .902,  .902}\textbf{ACN}} & \multicolumn{1}{c}{\textbf{KD-Net}} & \multicolumn{1}{c}{\cellcolor[rgb]{ .906,  .902,  .902}\textbf{ACN}} & \multicolumn{1}{c}{\textbf{KD-Net}} & \cellcolor[rgb]{ .906,  .902,  .902}\textbf{ACN} \\
    \midrule
          $\circ$     &$\circ$       &$\circ$     & \multicolumn{1}{c}{$\bullet$} & 39.04 & \cellcolor[rgb]{ .906,  .902,  .902}\textbf{42.98 } & 66.01 & \cellcolor[rgb]{ .906,  .902,  .902}\textbf{67.94 } & 82.32 & \cellcolor[rgb]{ .906,  .902,  .902}\textbf{85.55 } & 62.46  & \cellcolor[rgb]{ .906,  .902,  .902}\textbf{65.49 } \\
    \midrule
         $\circ$ &$\circ$       & \multicolumn{1}{c}{$\bullet$} & $\circ$      & 75.32 & \cellcolor[rgb]{ .906,  .902,  .902}\textbf{78.07 } & 81.89 & \cellcolor[rgb]{ .906,  .902,  .902}\textbf{84.18 } & 76.79 & \cellcolor[rgb]{ .906,  .902,  .902}\textbf{80.52 } & 78.00  & \cellcolor[rgb]{ .906,  .902,  .902}\textbf{80.92 } \\
    \midrule
          $\circ$& \multicolumn{1}{c}{$\bullet$} &$\circ$       &$\circ$       & 39.87 & \cellcolor[rgb]{ .906,  .902,  .902}\textbf{41.52 } & 70.02 & \cellcolor[rgb]{ .906,  .902,  .902}\textbf{71.18 } & 77.28 & \cellcolor[rgb]{ .906,  .902,  .902}\textbf{79.34 } & 62.39  & \cellcolor[rgb]{ .906,  .902,  .902}\textbf{64.01 } \\
    \midrule
    $\bullet$     &$\circ$       &$\circ$       &$\circ$       & 40.99 & \cellcolor[rgb]{ .906,  .902,  .902}\textbf{42.77 } & 65.97 & \cellcolor[rgb]{ .906,  .902,  .902}\textbf{67.72 } & 85.14 & \cellcolor[rgb]{ .906,  .902,  .902}\textbf{87.30 } & 64.03  & \cellcolor[rgb]{ .906,  .902,  .902}\textbf{65.93 } \\
    \bottomrule
    \end{tabular}%
  \label{compare_kdnet}%
\end{table}
\begin{figure*}
\includegraphics[scale=0.20]{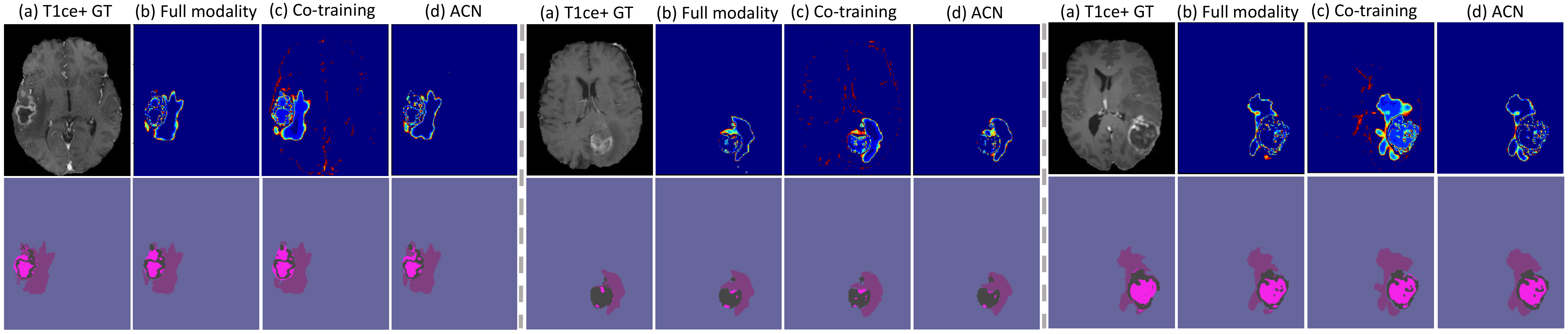}
   \caption{Qualitative results on BraTS2018 dataset. Column (a) shows one of the input modalities (T1ce) and the corresponding segmentation ground-truth. Column (b)(c)(d) show the prediction entropy maps from multimodal path (full modality), unimodal path (only T1ce modality) without ACN modules and unimodal path with ACN modules, along with their segmentation results.}
\label{fig:entropy_map}
\end{figure*}

\subsubsection{Ablation Study} In this section, we investigate the effectiveness of each proposed module, i.e., EnA, KnA and MMI. We choose a major modality T1ce as the single available modality. First we build a co-training baseline network which only uses consistency loss $\mathcal{L}_{\mathrm{con}}$. Then we add the three modules gradually. It is observed in Table \ref{tab:ablation} that the unimodal path for missing modality is improved simultaneously through adding these modules, which proves the mutual benefit of proposed co-training network and the cooperative work of the three modules.
Fig. \ref{fig:entropy_map} further verifies the superiority and rationality of the proposed modules. The entropy maps indicate the prediction of missing modality is less reliable than full modality, while our proposed ACN produces significantly better segmentation for missing modality with reduced uncertainty. 
\begin{table}[htbp]
  \centering
  \caption{Effectiveness of each module on unimodal (T1ce modality) paths (DSC \%).}
    \begin{tabular}{cccccccccc}
    \toprule
           &\multicolumn{1}{l}{$\mathcal{L}_{\mathrm{con}}$} & \multicolumn{1}{l}{EnA} & \multicolumn{1}{l}{KnA} & \multicolumn{1}{l}{MMI} & ET    & TC    & WT    & Average \\
    \midrule
    \multirow{5}[2]{*}{T1ce modality}  & $\surd$     &       &       &       & 74.57  & 82.08  & 77.72  & 78.12  \\
                 & $\surd$     & $\surd$     &       &       & 76.40  & 83.78  & 79.78  & 79.98  \\
                & $\surd$     & $\surd$     & $\surd$     &       & 77.14  & 83.85  & 78.82  & 79.94  \\
                 & $\surd$     & $\surd$     &       & $\surd$     & 76.12  & 84.24  & 79.89  & 80.09  \\
                 & \cellcolor[rgb]{ .906,  .902,  .902}$\surd$ & \cellcolor[rgb]{ .906,  .902,  .902}$\surd$ & \cellcolor[rgb]{  .906,  .902,  .902}$\surd$ & \cellcolor[rgb]{ .906,  .902,  .902}$\surd$ & \cellcolor[rgb]{ .906,  .902,  .902}\textbf{78.07 } & \cellcolor[rgb]{ .906,  .902,  .902}\textbf{84.18 } & \cellcolor[rgb]{ .906,  .902,  .902}\textbf{80.52 } & \cellcolor[rgb]{ .906,  .902,  .902}\textbf{80.92 } \\
    \bottomrule
    \end{tabular}%
  \label{tab:ablation}%
\end{table}%
\section{Conclusion}
In this work, we propose a novel Adversarial Co-training Network to address the problem of missing modalities in brain tumor segmentation. More importantly, we present two unsupervised adversarial learning modules to align domain and feature distributions between full modality and missing modality. We also introduce a modality-mutual information module to recover `missing' knowledge via knowledge transfer. Our model outperforms all existing methods on the multimodal BraTS2018 dataset in all missing situations by a considerable margin. Since our ‘One stop shop’ method needs to train 'dedicated' models for each missing situation, it may bring training cost, but it brings large improvement during inference without extra cost, which is of great value to clinical application and can be also generalized to other incomplete data domains.
%
%
%
 \bibliographystyle{splncs04}
 \bibliography{paper901}
\end{document}